\documentclass[10pt, conference]{IEEEtran}

\usepackage[colorlinks=true, linkcolor=blue, citecolor=blue,
  urlcolor=blue]{hyperref}

\usepackage{amsmath}
\usepackage{graphicx}
\usepackage{booktabs}
\usepackage{multirow}
\usepackage{tabularx}
\usepackage{enumitem}
\usepackage{threeparttable}
\usepackage{balance}
\usepackage{url}

\usepackage{fancyhdr} 
\fancypagestyle{firststyle}
{
\fancyhf{}
\fancyfoot[C]{\scriptsize{Proceedings of the IEEE International Conference on Software Analysis, Evolution and Reengineering (SANER 2026), Limassol, IEEE, pp. 1--6. \\ This version is the authors' copy. The publisher's definite version is available online via \url{https://doi.org/10.1109/SANER67736.2026.00118}.}}}

\begin{document}

\title{A Measurement Study on the Adoption of Pledges and Unveils in the
  OpenBSD Operating System}

\author{
\IEEEauthorblockN{Jukka Ruohonen}
\IEEEauthorblockA{University of Southern Denmark \\
Email: juk@mmmi.sdu.dk}
\and
\IEEEauthorblockN{Krzysztof Sierszecki}
\IEEEauthorblockA{University of Southern Denmark \\
Email: krzys@mmmi.sdu.dk}
\and
\IEEEauthorblockN{Abhishek Tiwari}
\IEEEauthorblockA{University of Southern Denmark \\
Email: abti@mmmi.sdu.dk}
}

\maketitle

\begin{abstract}
The paper presents a longitudinal measurement study on the adoption of the
\texttt{pledge} and \texttt{unveil} system calls in OpenBSD. These system calls
are used to sandbox programs and libraries. Given a dataset covering 19
releases, many programs and libraries were modified to use the system calls
already before their introductions in official releases. The adoption rates have
also steadily grown; a linear trend provides a coarse but sensible
heuristic. Although particularly programs residing in \texttt{/usr/bin} and
\texttt{/usr/sbin} have been modified to use the system calls, the sizes of
programs and libraries do not correlate well with the amounts of \texttt{pledge}
and \texttt{unveil} system calls invoked. Regarding the pledges made, standard
input and output operations have frequently been requested, although the full
fine-grained arsenal offered by \texttt{pledge} has generally been utilized in
OpenBSD. The same observation is seen in that particularly read operations to
given paths have frequently been unveiled. All in all, the measurement results
indicate that the adoption of system call minimization and sandboxing techniques
is not necessarily as troublesome as has often been discussed in the literature.
\end{abstract}


\section{Introduction}

\thispagestyle{firststyle} 

About a decade ago, the open source OpenBSD operating system introduced a new
\texttt{pledge} system call for minimizing attack surfaces related to system
calls. The \texttt{unveil} system call followed later. These two system calls
have received attention in the open source software world not only due to
security \textit{per~se} but also due to their simplicity and a different
implementation strategy compared to other related security
solutions~\text{\cite{Corbet22, Corbet18}}. However, academic attention has been
limited. In fact, it \textit{seems} that only two recent works have examined
the topic empirically, and these are based on interviews as well as a snapshot
of external packages distributed in OpenBSD and other open source operating
systems~\cite{Alhindi24, Alhindi25}. Thus, the present work continues the
earlier work by examining the longitudinal adoption of \texttt{pledge} and
\texttt{unveil} in OpenBSD itself, that is, the software explicitly developed
and maintained by OpenBSD developers themselves in their own source~code~tree. A
more detailed outlook on \texttt{pledge} and \texttt{unveil} is also provided.

The paper's overall motivation builds upon the arguments raised in the previous
work about difficulties to adopt system call minimization and sandboxing
techniques~\cite{Alhindi24}. A later study about the \texttt{seccomp} sandboxing
solution in Linux confirmed the many difficulties even experienced developers
have faced when trying to sandbox software~\cite{Alhindi25}. Although
difficulties and effort required are not explicitly observed in what follows,
the adoption rates observed can reveal whether the OpenBSD's sandboxing
solutions have been rolled throughout the operating system---despite potential
difficulties and obstacles. They can also indicate whether \texttt{pledge} and
\texttt{unveil} were originally designed well in terms of flexibility to adopt
them into different programs and libraries. In this regard, it has been argued
that \texttt{seccomp} is ``\textit{too} powerful'' \cite[p.~1]{Alhindi25} in a
sense that it provides a whole sandboxing toolbox. In contrast and true to the
Unix philosophy~\cite{Raymond03}, the \texttt{pledge} and \texttt{unveil} system
calls are rather simple to use---at least on paper.

Regarding the remaining structure, the background is first motivated in the
opening Section~\ref{sec: background}. It also introduces four research
questions for the empirical measurement analysis. Data and methods are presented
in the subsequent Section~\ref{sec: materials and methods} after which results
follow in Section~\ref{sec: results}. The last three Sections~\ref{sec:
  conclusion}, \ref{sec: limitations}, and \ref{sec: closing remarks} present
the conclusion reached, a couple of limitations, and two concluding remarks,
respectively.

\section{Background}\label{sec: background}

\subsection{System Call Minimization}\label{subsec: system call minimization}

Attack surface minimization is a classical security engineering technique. It is
also required by a recent European regulation, and the minimization techniques
include everything from disabling or removing unnecessary features to reducing
wireless signal propagation~\cite{Mansori23, Ruohonen25ESPREa}. Given a lack of
universally accepted definitions~\cite{Theisen18}, in what follows, the terms
attack surface and attack surface minimization are restricted to system calls in
operating systems and their reduction for userland processes. Despite the
restriction, the conventional logic applies; the more there are unnecessary
system calls available for a running process, the larger the attack surface of
an operating system via the process. In addition to bluntly disabling system
calls within a kernel itself, sandboxing userland processes is a common
technique, and, therefore, also the \texttt{pledge} and \texttt{unveil} system
calls have often been labeled as sandboxing techniques~\cite{Alhindi24,
  Anderson17}. Fundamentally, the issue is that operating systems and their
kernels offer too much functionality, and thus developers using the sandboxing
techniques seek to ``prevent as much as possible'', to quote an interviewee's
comment from an existing study~\cite[p.~4]{Alhindi25}.

The idea about minimizing attack surfaces through system call minimization is
not new~\cite{Bernaschi00}. It has also continuously been adopted to new
contexts, including binaries~\cite{Mansori23} and virtualization techniques
\cite{Ghavamnia20, Yang24, Zhan23}. Also OpenBSD has a long tradition in system
call minimization implementations.

Regarding history and the OpenBSD's software evolution, the policy-based
\texttt{systrace} utility from the early 2000s~\cite{Provos03} was deprecated in
2016 from OpenBSD with the introduction of \texttt{pledge}. The \texttt{pledge}
and \texttt{unveil} system calls are also used directly in a source code. Their
underlying idea is close to the ideas of the \texttt{seccomp} system call
available in Linux~\cite{Bechis17} and the capability-based \texttt{capsicum}
implementation in FreeBSD~\text{\cite{FreeBSD24capsicum, Watson10}}. To use
OpenBSD's parlance, a software developer thus ``promises'' with \texttt{pledge}
to use only certain system calls and their features. For instance, the old
\texttt{hack} game, which is about ``exploring the dungeons of
doom''~\cite{OpenBSD25hack}, promises only to use standard input and output:

\begin{verbatim}
     if (pledge("stdio", NULL) == -1)
             err(1, "pledge");
\end{verbatim}

If the given process annotated with \texttt{pledge} does not keep its promise
while running, the kernel will kill and report it. Implicitly and theoretically,
a similar idea can be seen to work underneath different assertions used to
verify preconditions and the future behavior of
software~\cite{Ruohonen25ICTSSa}. The \texttt{unveil} system call complements
\texttt{pledge}. When using the system call, a developer seeks to limit access
to a file system; each and every attempt to access a path not explicitly
unveiled returns an error.

\subsection{Benefits}

The \texttt{pledge} and \texttt{unveil} system calls have many security benefits
because they allow fine-grained restrictions at runtime. For instance, both
dropping privileges and privilege separation benefit from the use of the
\texttt{pledge} system call~\text{\cite{Beck18, DeRaadt16}}. (The former means
that a process that must start with \texttt{root} privileges drops its
privileges to those of a normal, restricted user later on, whereas the latter
means that a process that must run with \texttt{root} privileges does so only
minimally, outsourcing many functions to other child processes and their own
users.) Also restricting access to a file system comes with a straightforward
rationale. Given a ``virtue to simplicity in security'', a ``compromised process
cannot exfiltrate data that it cannot read, and it cannot corrupt files that it
cannot write''~\cite{Corbet18}. Thus, the overall rationale of \texttt{unveil}
can be compared to other file system integrity solutions, such as the
policy-based \texttt{veriexec} subsystem in NetBSD~\cite{NetBSD17veriexec}. The
difference again is that the \texttt{unveil} call is used directly in a
source~code.

\subsection{Adoption Challenges}\label{subsec: challenges}

Yet, there are also practical challenges. For various reasons, including
knowledge gaps and effort required, there are well-known and enduring challenges
in the adoption of software security techniques and
features~\cite{Hermann25}. Although the situation has improved over the years,
particularly the adoption of proactive, preventive security engineering
techniques and features has been observed to be slower than the adoption of
reactive techniques, including handling of software
vulnerabilities~\cite{Weir21}. A potential explanation may relate to
observations according to which much of security engineering activities tend to
focus on post-development verification before product
launches~\cite{Rindell21IST}. Therefore, it is not surprising that criticism has
been expressed also about sandboxing techniques, including with respect to
\texttt{pledge}, \texttt{seccomp}, and \texttt{unveil}
specifically~\text{\cite{Alhindi24, Yang24, Zhan23}}. Despite research,
automation remains a challenge~\cite{Yang24}, among other things. Similar
criticism applies to policy-based solutions, such as the Security-Enhanced Linux
(SELinux) implementation~\cite{Eaman17}. Particularly in case other developers
need to do the work, it is supposedly difficult to get them to adopt the
security techniques. Debugging is hard and time-consuming. That said, it should
be noted that similar points apply also more generally; the adoption of new
application programming interfaces has been observed to be
slow~\cite{McDonnell13}, for instance.

These potential adoption challenges have been implicitly acknowledged by OpenBSD
developers who have listed various debugging tools for helping developers to
adopt \texttt{pledge} and \texttt{unveil}~\cite{Bechis17, DeRaadt16}. That said,
the BSD operating systems have an edge in this regard because they are complete
operating systems, meaning that they have their own kernels and non-kernel
userlands. Therefore, the adoption challenges are presumably more on the side of
the external software available via the OpenBSD's package manager. In this
regard, criticism has been expressed that adoption may be difficult particularly
for large and complex software~\cite{Anderson17}. As the paper is restricted to
the OpenBSD's evolving userland, these points about potential adoption
challenges are not examined to the full extent possible. Nevertheless, already a
case study, which is about studying a~``contemporary phenomena in its natural
context''~\cite[p.~131]{Runeson09}, helps at understanding how well
\texttt{pledge} and \texttt{unveil} have been adopted in their natural context,
that is, the OpenBSD operating system. To the best of the authors' knowledge,
the paper is also the first to empirically study the longitudinal adoption of
the system~calls.

\subsection{Research Questions}

The preceding discussion motives the following four research questions (RQs) for
guiding the empirical analysis:
\begin{itemize}
\itemsep 3pt
\item{RQ.1: What have been the adoption rates of \texttt{pledge} and
  \texttt{unveil} in the OpenBSD's recent history?}
\item{RQ.2: Has software size affected the adoption rates?}
\item{RQ.3: What have been frequently promised with \texttt{pledge} and unveiled with \texttt{unveil}?}
\item{RQ.4: Irrespective of the answer to RQ.1, have the promises and unveils
  between stable across releases?}
\end{itemize}

With these research questions, the paper further interlaces with the large
empirical software evolution research branch. In this regard, the Linux kernel
has been a popular choice for case studies~\cite{Ruohonen25ICTSSa}. Although
there are case studies and comparative studies about BSD operating
systems~\text{\cite{Ruohonen15IWPSE, Ruohonen19RSDA, Spinellis21, Zhao25}},
including OpenBSD specifically~\cite{Alhindi24, Shi23}, their amount seems
limited when compared to those addressing Linux and open source software in
general. Therefore, a case study on OpenBSD's new security features also makes a
small contribution to empirical software engineering knowledge in~general.

\section{Materials and Methods}\label{sec: materials and methods}

\subsection{Data}\label{subsec: data}

The \texttt{pledge} system call first appeared in OpenBSD~5.9, whereas
\texttt{unveil} was introduced a little later, in the 6.4
release~\cite{OpenBSD25pledge, OpenBSD25unveil}. Therefore, the dataset, which
is also available online for replication purposes~\cite{Ruohonen25DSSANER}, is
based on OpenBSD releases from 5.9 to 7.7. The period covered is from March 2016
to April 2025. The 7.7 release was the latest available at the time of
writing. In total, $19$ releases are observed. Although the amount is a
borderline case for using formal time series methods, the calendar time span of
about a decade is more than sufficient for answering to RQ.1 and RQ.4.

The archived (\texttt{src.tar.gz}) source code trees were retrieved from a
mirror archiving the OpenBSD's full
history.\footnote{~\url{https://mirror.leaseweb.com/pub/OpenBSD/}} By following
a parsing strategy used in related work~\cite{Ruohonen25ICTSSa}, the dataset
was constructed by processing each file ending to a suffix \texttt{.c} line by
line, identifying promises and unveils through occurrences of the
\texttt{\nonfrenchspacing pledge} and \texttt{unveil} character strings followed
by an opening parenthesis. The \texttt{distrib}, \texttt{gnu}, \texttt{regress},
and \texttt{sys} root directories in the OpenBSD's source code tree were
excluded from the parsing. These directories contain bundled software shipped
with a GNU license, regression tests, kernel code, and miscellaneous~files.

\subsection{Metrics and Measurements}\label{subsec: metrics}

The measurement analysis operates with counts of calls to the two system calls
as well as counts of the parameters to these calls. Regarding the latter counts,
the measurements focus on the first parameters supplied to \texttt{pledge} and
the second parameters supplied to \texttt{unveil}. These are elaborated further
when presenting the results in Subsection~\ref{subsec: promises and
  unveils}. For the time being, it suffices to only note that the second
parameters to \texttt{pledge} were omitted as almost all were \texttt{NULL}
values and that the second parameters to \texttt{unveil} concern permissions
granted to paths supplied as the first parameters; the paths themselves are not
interesting to measure without a context.

The directories in the source code tree mimic the directories of a deployed
OpenBSD system. For instance, the directory \texttt{usr.sbin} contains programs
that will be installed into the \texttt{/usr/sbin} directory during a
deployment. Programs and libraries are identified through sub-directories;
hence, \texttt{usr.sbin/cron} is about the \texttt{cron} scheduling daemon. It
should be remarked that the operationalization is not perfect because some
sub-directories may contain files that will be installed as multiple executable
binaries or shared libraries. While from a developers' perspective these can be
argued to still belong to a single unifying ``program'', the terminological
choice should be kept in mind when interpreting the results.

Regarding methods, descriptive statistics are sufficient for answering to the
four RQs---keeping also in mind the note in Subsection~\ref{subsec: data} about
the sample size. Regarding RQ.2, correlations are reported. The rationale comes
from a phrasing that \texttt{pledge} and \texttt{unveil} ``are designed to make
big programs safer''~\cite{Beck18}. Thus, a \textit{hypothesis} is that these
have been enrolled particularly for larger programs and libraries during the
past decade of the OpenBSD's evolution. Given the remark about software sizes
and complexity made earlier in Subsection~\ref{subsec: challenges}, it should be
remarked that the programs and libraries in the OpenBSD's userland are still
rather small when compared to, say, web browsers or desktop
environments. A~related point is that OpenBSD contains also fewer system calls
and default userland software than other operating systems, including FreeBSD,
macOS, and Ubuntu~\cite{Spinellis21}. Thus, it can be also argued that the
OpenBSD's overall attack surface is already likely smaller than in many other
operating systems.

\section{Results}\label{sec: results}

\subsection{Adoption Rates (RQ.1)}

\begin{figure}[th!b]
\centering
\includegraphics[width=\linewidth, height=4cm]{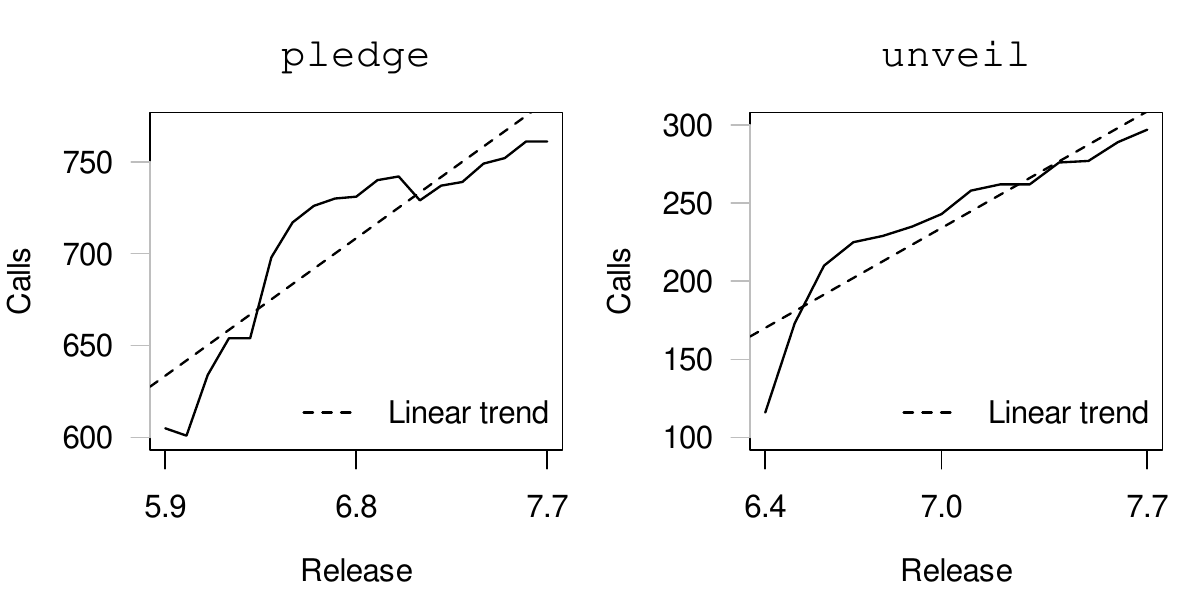}
\caption{Calls to \texttt{pledge} and \texttt{unveil} across releases}
\label{fig: calls}
\end{figure}

The adoption rates of the two system calls can be deduced already from
Fig.~\ref{fig: calls}, which shows the counts of calls to \texttt{pledge} and
\texttt{unveil} across the nineteen releases observed. When starting the
interpretation from the $y$-axes of the two plots, it becomes clear that a
substantial amount of software was adopted to use the system calls already prior
to the introduction of the two system calls in formal releases. After their
introductions, the adoption rates have continued to grow. While not perfectly,
as curve fitting seldom does, a linear trend seems to provide sensible enough
estimates for the growth rates. The latest 7.7 release observed already contains
over $700$ calls to \texttt{pledge} and nearly $300$ calls to
\texttt{unveil}. All in all, these observations align well with earlier results
according to which in 2023 over 90\% of all running processes were sandboxed in
a fresh install of OpenBSD~\cite{Alhindi24}. At the time of writing, in late
2025, the share could be even slightly~higher.

\begin{figure}[th!b]
\centering
\includegraphics[width=\linewidth, height=4cm]{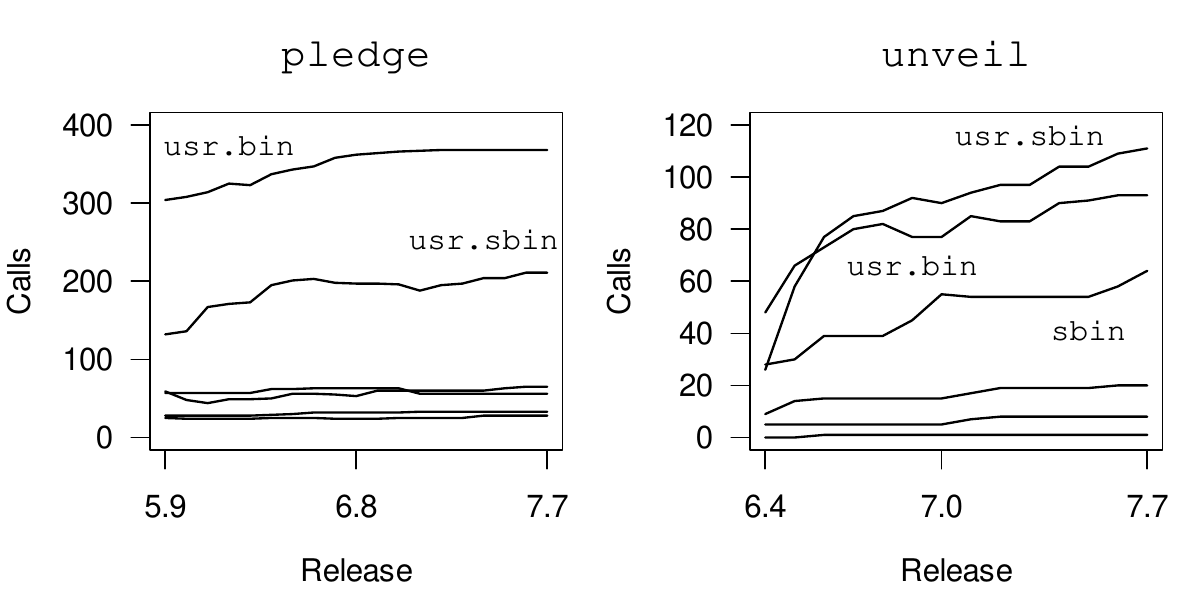}
\caption{Calls to \texttt{pledge} and \texttt{unveil} across root directories}
\label{fig: dirs}
\end{figure}

As can be seen from Fig.~\ref{fig: dirs}, many of the calls are within programs
distributed within the \texttt{/usr/bin} and \texttt{/usr/sbin} directories in a
deployed OpenBSD system. Although also \texttt{/sbin} stands out in terms of
\texttt{unveil} calls, the two directories under \texttt{/usr} are not primarily
about fundamental yet simple programs. Therefore, an argument about
\texttt{pledge} having been explicitly or implicitly designed for ``trivial
OpenBSD base system applications'' \text{\cite[p.~22]{Anderson17}} does not hold
ground.

\subsection{Software Size (RQ.2)}

The top-25 programs and libraries in terms of invocations of the two system
calls are shown in Fig.~\ref{fig: calls}. Regarding \texttt{pledge}, OpenSSL
leads the ranking. The explanation traces to the operationalization based on
sub-directories (see Subsection~\ref{subsec: metrics}). That is: the
\texttt{usr.bin/openssl} directory contains many small programs, most of which
have a single \texttt{pledge} call in their \texttt{main} functions. The second
place is taken by \texttt{smtpd}, a Simple Mail Transfer Protocol (SMTP) daemon
developed by OpenBSD. Thereafter, the call counts decrease rapidly, eventually
reaching the median of one call across the median of calls made across the
nineteen releases. Regarding \texttt{unveil}, many fundamental programs, such as
\texttt{su} and \texttt{passwd}, have been refactored to limit file system
access. Of the larger programs, Dynamic Host Configuration Protocol (DHCP) and
Lightweight Directory Access Protocol (LDAP) clients, \texttt{dhcpleased} and
\texttt{ldapd}, are worth pointing out as examples of larger programs making
many \texttt{unveil} calls.

\begin{figure}[th!b]
\centering
\includegraphics[width=\linewidth, height=8cm]{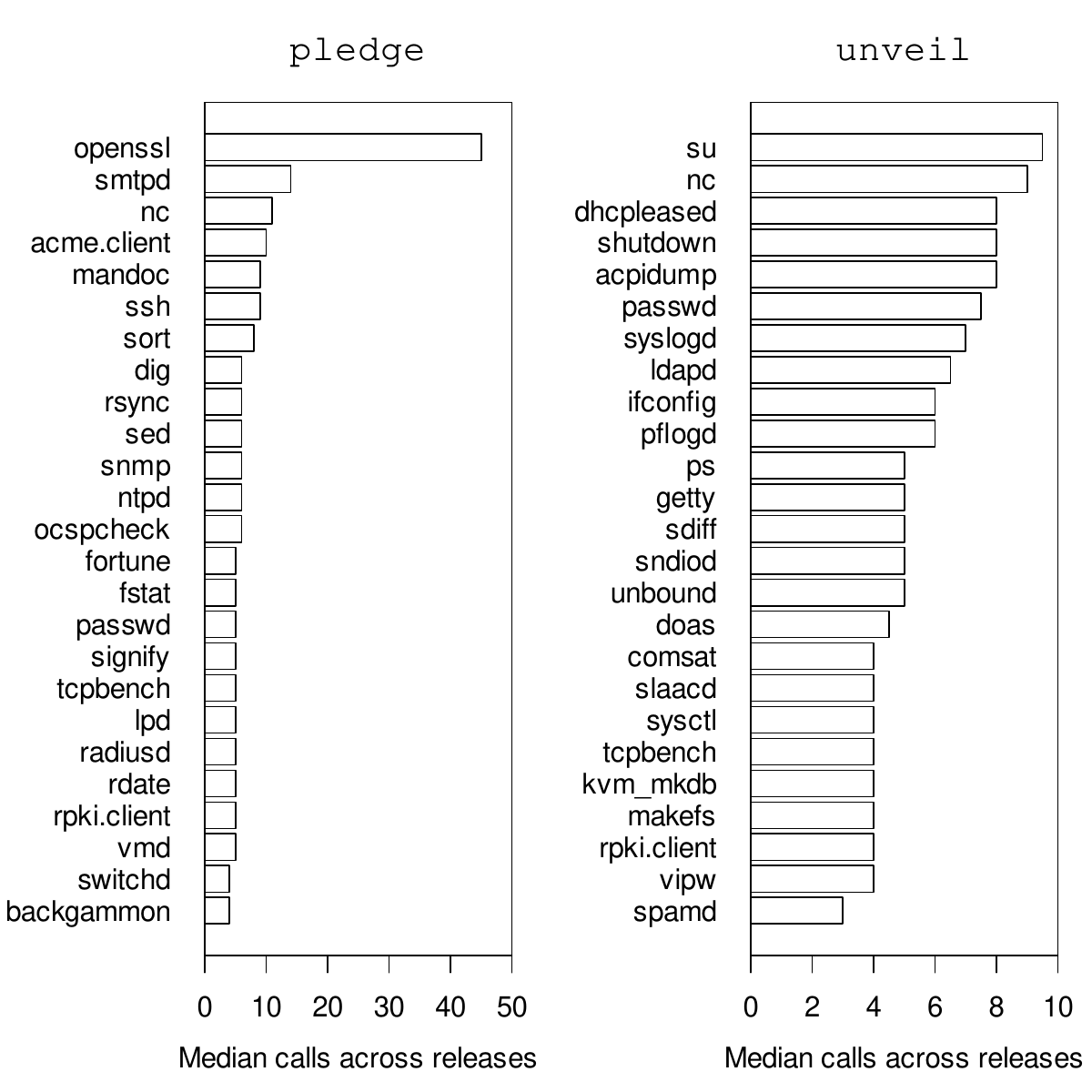}
\caption{Top-25 programs and libraries using \texttt{pledge} and \texttt{unveil}}
\label{fig: programs}
\end{figure}

\begin{figure}[th!b]
\centering
\includegraphics[width=\linewidth, height=4cm]{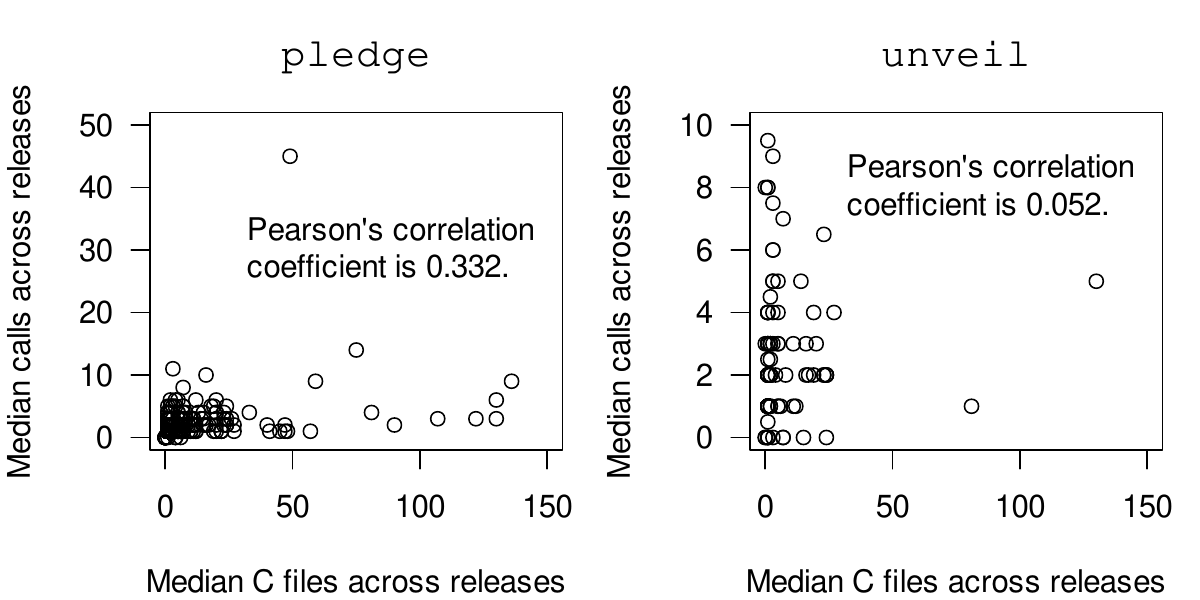}
\caption{Correlations between calls and C files}
\label{fig: cor}
\end{figure}

However, the sizes of programs and libraries do not correlate well with the
system call counts. When using the number of files ending to a \texttt{.c}
suffix within a given sub-directory under the root directories in Fig.~\ref{fig:
  dirs} as a proxy for software size, as has been done also in previous
research~\cite{Ruohonen25ICTSSa}, a visualization in Fig.~\ref{fig: cor} makes
the observation clear. For \texttt{unveil}, the correlation coefficient is tiny
($0.052$). Although a higher value is present for \texttt{pledge}, also it is
small in magnitude~($0.332$). When using mean instead of median, the correlation
coefficients are even lower in their magnitudes. OpenSSL, which is denoted by
the topmost circle on the left-hand side plot of Fig.~\ref{fig: cor}, is among
the exceptions. All in all, it cannot be said that larger software would have
been modified to use the system calls more than smaller software or the other
way~around.

\subsection{Promises and Unveils (RQ.3 and RQ.4)}\label{subsec: promises and unveils}

\begin{figure}[th!b]
\centering
\includegraphics[width=\linewidth, height=8cm]{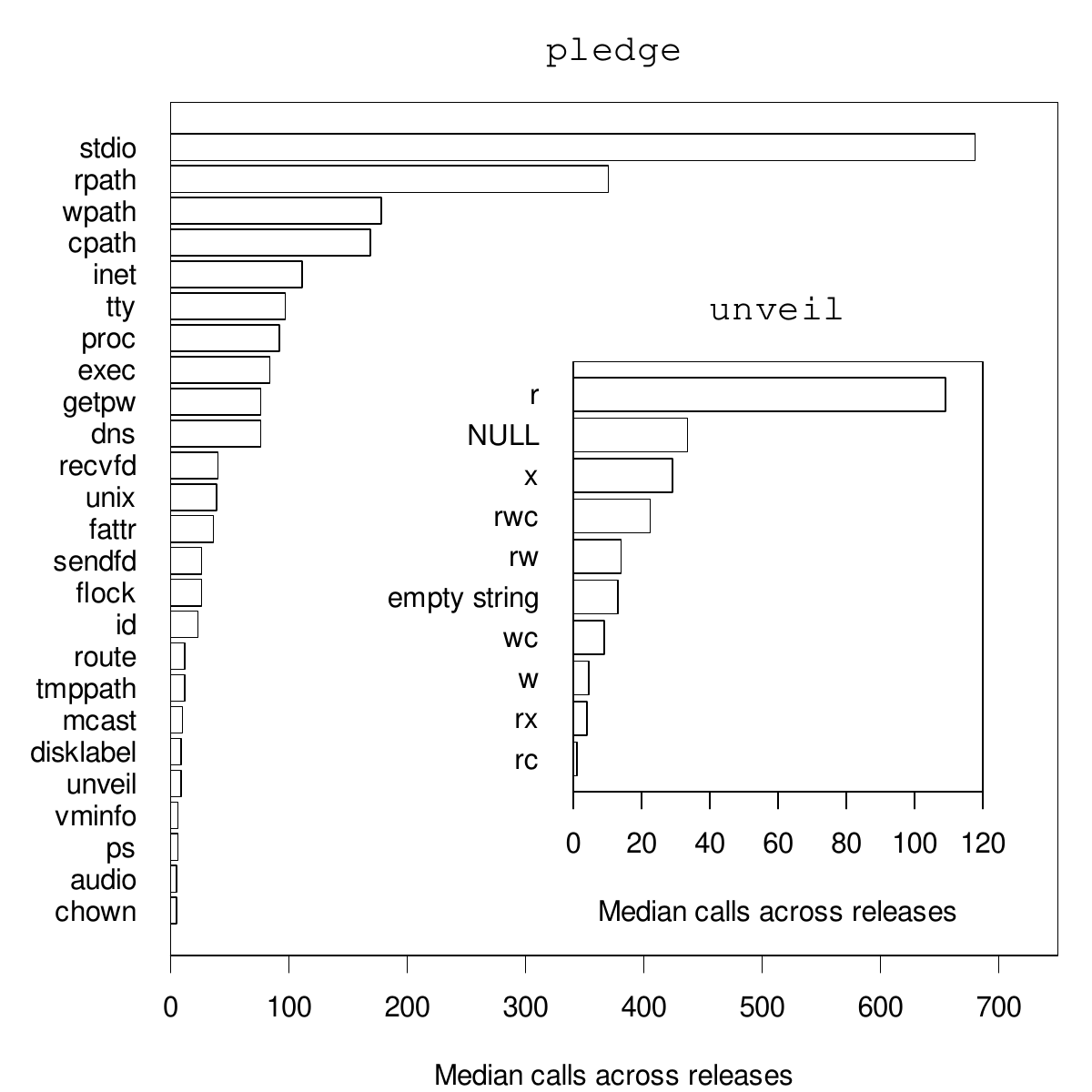}
\caption{Top-25 promises to \texttt{pledge} and all unveils to \texttt{unveil}}
\label{fig: flags}
\end{figure}

The top-25 most common promises (outer plot) and unveils (inner plot) are shown
in Fig.~\ref{fig: flags}. Regarding the promises, which, to recall from
Subsection~\ref{subsec: metrics}, are about the first parameters supplied to
\texttt{pledge}, standard input and output operations available via the standard
C library have frequently been promised and thus used. The observation is hardly
surprising, given the notion that in Unix-like operating systems ``everything is
a file'' (or a file descriptor)~\cite{Brown10}. The second, third, and fourth
places are taken by the \texttt{rpath}, \texttt{wpath}, and \texttt{cpath}
parameters, which too are about input and output operations. Thereafter, the
picture starts to become more nuanced, including promises about sockets
(\texttt{inet}), terminals (\texttt{tty}), processes (\texttt{proc} and
\texttt{execve}), and many others.

\begin{figure}[th!b]
\centering
\includegraphics[width=\linewidth, height=4cm]{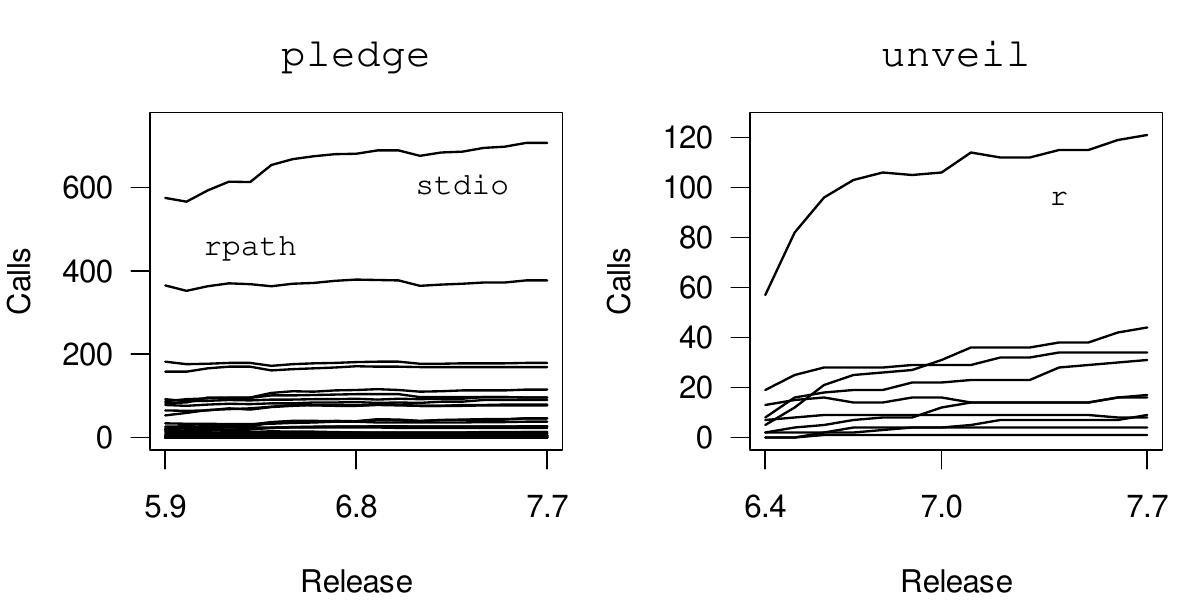}
\caption{Promises to \texttt{pledge} and unveils to \texttt{unveil} across releases}
\label{fig: flags releases}
\end{figure}

Regarding the unveils, over 40\% of the calls to \texttt{unveil} have used the
\texttt{r} permission for requesting paths to be readable. These calls are
similar to \texttt{pledge} calls with the \texttt{rpath} promise. The other
unveils requested are far less common. Furthermore, the counts of the promises
and unveils have been rather stable across the releases, as seen from
Fig.~\ref{fig: flags releases}. Only the \texttt{stdio} and \texttt{r}
parameters have seen visible growth, thus correlating slightly with the earlier
trends in Fig.~\ref{fig: calls} and Fig.~\ref{fig: dirs}.

\section{Conclusion}\label{sec: conclusion}

The paper presented a longitudinal measurement study about the adoption rates of
the \texttt{pledge} and \texttt{unveil} system calls for system call
minimization and sandboxing in general. The primary conclusion is clear: the
system calls have been widely and continuously adopted in the OpenBSD's core
userland. Both programs and libraries have been modified to use the system
calls. The sizes of programs and libraries do not correlate well with the
adoption rates; it cannot be said that smaller or larger programs would have
seen stronger or weaker adoption. Rather, \texttt{pledge} and \texttt{unveil}
have seen wholesale adoption. Thus, the earlier contradictory quotations about
``big programs''~\cite{Beck18} and ``trivial'' programs~\cite{Anderson17} both
seem incorrect for describing the adoption rates; the two system calls work for
both types of programs, although their future success remains unclear at this
point. Even though input and output operations have frequently been requested
via the system calls, the full arsenal offered has generally been used. This
point reiterates the wholesale adoption, but it can be interpreted to also
indicate adaptability and flexibility of the two system calls. That said,
maintenance effort required remains unclear. Given existing results about
maintenance problems with some specific system calls, such as \texttt{ptrace} in
Linux~\cite{Bagherzadeh18}, the alleged ``maintenance nightmare'' with
sandboxing techniques~\cite{Alhindi25} would offer a good topic for further
longitudinal software measurement research.

\section{Limitations}\label{sec: limitations}

Generalizability is always a limitation with case studies. However, a case study
seems as the only plausible option to examine the topic at hand from a
quantitative software metrics perspective. Given the framing that was done in
Subsection~\ref{subsec: system call minimization} about attack surfaces, it
seems \texttt{seccomp} and the FreeBSD's \texttt{capsicum} would be the only
plausible options for enlarging a sample toward other operating
systems. However, direct comparisons would be difficult due to different
functionalities. (While there have been talks and attempts to port
\texttt{pledge} to Linux~\cite{Corbet22}, their success remains unclear at the
time of~writing.) Another option would be to continue the noted earlier
work~\cite{Alhindi24} by examining external packages.

Another limitation originates from potential parsing errors. In particular, the
promises for \texttt{pledge} were initially taken from the online
documentation~\cite{OpenBSD25pledge}, and then augmented with five additional
promises that were not documented. As others were excluded, it is possible that
some promises are missing. That said, the potential effect of these potential
exclusions upon the answers to RQ.3 and RQ.4 seem rather minimal.

A final noteworthy limitation concerns operationalization. Among other things,
the crude proxying of software size via C files could be augmented with more
elaborate metrics, including those measuring complexity explicitly. Given the
discussion in Subsection~\ref{subsec: challenges}, code churn, commit
reverts~\cite{Yan19}, and related aspects could reveal insights about potential
challenges despite the continuous adoption rates observed. Regarding software
evolution, the adoption rates should also be scaled by growing size of the
operating system itself. In other words, many---if not all---operating systems
are continuously growing and becoming more complex~\cite{Zhao25}, and complexity
is known to impede the adoption of security features and
practices~\cite{LariosVargas23}.

\section{Closing Remarks}\label{sec: closing remarks}

Two points can be raised as more general and constructively critical
takeaways. The first point is about the general adoption challenges discussed in
Subsection~\ref{subsec: challenges}. Given the proliferation of minimization
techniques and sandbox implementations across open source operating
systems~\text{\cite{Alhindi24, Anderson17}}, it seems sensible to raise a
critical point that a unified standard might bring wider benefits throughout the
open source software world. The Portable Operating System Interface (POSIX) set
of standards might be something to consider in this regard. Given the strong
adoption rates observed and reported, maybe there might be something to learn
from OpenBSD and its developers too?

The second point is about measuring software security. Although it has been
observed that vulnerabilities have reduced in OpenBSD over time, it remains
debatable whether a conclusion about improved security~\cite{Shi23} can be done
by observing vulnerabilities~\cite{CS25, Meneely25}. Alongside decreased
vulnerability counts, OpenBSD has introduced numerous exploitation mitigation
techniques, including various randomization solutions. Against this backdrop, it
seems reasonable to end the paper with an argument regarding the continuing
debate about measuring software \text{security---also} the often overlooked
adoption of security techniques and features should be measured. It would be
also worthwhile to try to measure whether the continuous adoption rates of
\texttt{pledge} and \texttt{unveil} have correlated with a potentially
decreasing amount of exploits available. Finally, it should be emphasized that
measurements alone are not sufficient; verification research is needed too.

\balance
\bibliographystyle{IEEEtran}


\end{document}